%% file: draft.tex
\documentstyle[osa,epsfig]{revtex}
\begin{document}            
\sloppy         
\title{A Monte Carlo Study of Equilibrium Polymers in a Shear Flow} 
\author{A. Milchev$^1$(\footnote{Email for communication:
milchev@ipchp.ipc.bas.bg or j.wittmer@dpm.univ-lyon1.fr})
J. P. Wittmer$^2$, and D. P. Landau$^3$}
\vskip 1.0truecm
\address{
$^1$ Institute for Physical Chemistry, Bulgarian Academy of Sciences, $1113$
Sofia, Bulgaria\\
$^2$ Departement de Physique des Materiaux, Universit\'e Claud Bernard
Lyon I,\\ 
69622 Villeurbanne, CEDEX, France\\
$^3$ Department of Physics and Astronomy, University of Georgia,
Athens, Ga. 30602, USA}
\maketitle
\begin{abstract}            
We use an off-lattice microscopic model for solutions of equilibrium 
polymers (EP) in a lamellar shear flow generated by means of a self-consistent 
external field between parallel hard walls. 

The individual conformations of the chains are found to 
elongate in flow direction and shrink perpendicular to it while
the average polymer length decreases with increasing
shear rate.

The Molecular Weight Distribution of the chain lengths retains largely its 
exponential form in dense solutions whereas in dilute solutions it
changes from a power-exponential Schwartz distribution to a purely 
exponential one upon an increase of the shear rate.

With growing shear rate the system becomes increasingly inhomogeneous
so that a characteristic variation of the total monomer density, the 
diffusion coefficient, and the center-of-mass distribution of polymer 
chains of different contour length with the velocity of flow is observed.

At {\em higher} temperature, as the average chain length decreases significantly, 
the system is shown to undergo an order-disorder
transition into a state of nematic liquid crystalline order with an easy
direction parallel to the hard walls. The influence of shear flow on this
state is briefly examined. 
\end{abstract} 
\vskip 1.0truecm
\today
\input{intro}
\input{model}
\input{bias}
\input{end}

\input{bib}
\end{document}

%% file: intro.tex
\section{Introduction}

Systems in which polymerization takes place under condition of chemical 
equilibrium between polymer chains and their respective monomers are 
termed "equilibrium polymers" (EP)\cite{ref}.
The interest to EP from the point of view of both applications
and basic research has recently triggered numerous investigation,
including computer simulations\cite{WMC2,K} in an effort to avoid difficulties with laboratory 
experiments\cite{Greer} and approximations as the Mean Field Approximation (MFA).
Recently the basic scaling concepts of polymer physics were tested by 
extensive Monte Carlo (MC) simulations of flexible EP on a lattice\cite{WMC2}.
The results suggest that despite polydispersity, EP resemble conventional
polymers (where the polymerization reaction has been deliberately terminated) in many 
aspects. 
However, dynamic aspects of their behavior may still be very 
different: for example, the constant process of scission and recombination 
in EP offers an additional mechanism of stress relaxation\cite{Cates}. 
Computer experiments on EP dynamics are already under 
way\cite{MWL1}.  

Considerably fewer simulation studies of {\em non-equilibrium} properties of EP 
have been reported\cite{K,AMYRDL}. Recently observed phenomena such 
as shear banding structure, shear inducing structure and phase 
transitions\cite{Berret,Schmitt,Makhloufi,Furo} are not completely understood. 
An earlier theoretical work\cite{Gelbart}, for instance, predicted a decrease in 
average size of dilute rod-like micelles whereas a later study\cite{Wang} 
concluded that rod-like micelles should grow at higher shear rates. Since it 
is known that viscoelastic surfactant solutions show unusual nonlinear rheology%
\cite{Spenley}, it is clear that much more research in this field is 
needed before complete understanding of nonlinear properties of EP is
achieved. 

Since EP behave in many respects as conventional "dead" polymers%
\cite{WMC2}, comparisons with the latter where much more work on shear 
flow effects has been done so far, could prove very useful. Thus
inhomogeneity of flows, due to the presence of boundaries, and its
impact on polymer behavior may be directly observed experimentally
by means of evanescent wave-induced fluorescence method\cite{Rondel} 
that can probe the polymer concentration in the depletion layer adjacent 
to the walls. Coil stretching of dilute flexible polymers in a flow,
diffusion and density profiles as well as "slip" effects near walls
have been treated theoretically\cite{deG,Aubert,Onuki,Biller} and 
by computer simulations\cite{Goh,Diaz,Duering}, and as we shall demonstrate 
below, many of these early results compare favorably with what we
observe for EP in the present investigation.

In the present study we employ a dynamic Monte Carlo algorithm in order
to study EP properties in shear rate. The flow of the system in  a
semi-infinite slit of thickness $D$ is induced by applying an external
field $F$ with magnitude which changes linearly across the slit and is
parallel to the hard walls of the container. Thus the jump rate of the monomers
becomes thus biased along the $x$-axis and a flow of the system through
the periodic boundary sets in. 

One should emphasize that such an investigation
should focus on the linear response in a laminar shear flow. 
MC methods cannot account for
hydrodynamic interactions in principle and the transition from laminar 
to turbulent flow can be simulated by means of Molecular Dynamics (MD) only.
The linear response breaks down at field intensities when the maximum flow 
velocity is attained, i.e. when all $100 \%$ of
the random jumps along the $x$-axis are forced to occur, say, in positive
direction. Any further increase of the field $F$ will then fail to 
accelerate the particles any further. Even with these limitations,
however, is appears that this kind of MC simulation of EP in a shear flow is
warranted, given the considerably longer time periods or systems sizes
a MC methods may handle as compared to MD.

All Monte Carlo studies of EP so far have been performed
on a cubic lattice either exploiting an analogy of the Potts model of
magnetism to random self-avoiding walks on a lattice\cite{MilchevPotts,ML}, 
or using the Bond Fluctuation (BFL) Model\cite{WMC2,AMYRDL}. These lattice
models were developed and 
extensively used for monodisperse systems of conventional polymers and are 
known to faithfully reproduce their dynamic (Rouse) behavior. 
For the purpose of shear flow studies a disadvantage of these models , due 
to the discrete structure of the lattice, appears obvious: monomers would 
block each other on the lattice at higher shear rates. Random jumps would have to 
be of the size of single monomers only, and, last not least, the artificial 
cubic symmetry would predetermine ordering effects along the three major 
axes of the lattice\cite{WPB} thereby questioning  possible phase transitions
into liquid crystalline order.

In the present work we employ an off-lattice model of EP, designed to
overcome these and other shortcomings of previous lattice models and to serve
in examining the role of polymers (semi)-flexibility. An off-lattice model should
be a better tool in dynamic studies of a broader class of soft condensed
matter systems where bifunctionality of the chemical bonds might be
extended to polyfunctional bonds, as this is the case in gels and membranes.
A comprehensive comparison of this off-lattice algorithm to earlier
lattice models\cite{WMC2} shows that all properties of EP
derived in former investigations, are faithfully reproduced
in the continuum too. 

%% file: model.tex
\section {Description of the model}
\label{model}

As in our earlier off-lattice bead-spring model of conventional polymer 
chain\cite{Gerroff,MPB}, a coarse-grained polymer chain consists of $l$ beads 
or "effective monomers". These are connected by springs which represent "effective 
bonds" and are described by a FENE (finitely extendible non-linear elastic) 
potential:
\begin{eqnarray}  \label{FENE}
U_{FENE}(r) & = & -\frac{k}{2}R^2\log \left[1-\left(\frac{r-r_0}{R}\right)^2
\right] - J,\quad \ \mbox{for}\ 
\quad -R<r-r_0<R, \\
U_{FENE}(r) & = & \infty ,\ otherwise\,  \nonumber
\end{eqnarray}
where $r$ is the distance between two successive beads, $r_0 = 0.7$ is the
unperturbed bond length with maximal extension $l_{max}$, $R = l_{max}-r_0 
= 0.3$, and $k/2=20$ (in our units of energy $k_BT = 1.0$) is the elastic 
constant of the FENE potential which behaves as a harmonic potential for 
$r-r_0 \ll R$. Thus $U_{FENE}(r \approx r_0) \approx
-\frac{k}{2}(r-r_0)^2$ but diverges logarithmically both for $r \rightarrow
l_{max}$ and $r \rightarrow l_{min} = 2r_0 - l_{max}$. We choose our unit
of length such that $l_{max} = 1$ and then the hard core diameter of the
beads $l_{min} = 0.4$. All lengths as, e.g. the linear size of the 
simulational box, are then measured in units of $l_{max}$. 

According to Eq.(\ref{FENE}) the net gain of energy of a monomer which forms a 
bond with a nearest neighbor at 
distance $r_0$ is then equal to the "bond" energy $J$.
In EP these strong attractive bonds between nearest neighbors
along the backbone of a chain are constantly subject to scission and
recombination. In the present model only bonds, stretched a distance $r$ 
beyond some threshold value, $r_{break}=0.8l_{max}$,  attempt to 
break so that
eventually an energy $U_{FENE}(r) > 0$ in the interval between $0$ and $J$
could be released if the bond is broken.

Each monomer has two unsaturated bonds which may be either engaged in forming
a strong saturated bond between nearest neighbors along the backbone of a chain
(when the originally unsaturated bonds of such neighbors meet and become a 
parallel pair) or remain free (or "dangling") as in the case of chain
ends or non-bonded single monomers. In order to create a bond, however,
the respective monomers must  approach
\footnote{Recombination for $r < r_{break}$ would violate detailed balance 
if scissions occur at $r > r_{break}$ only.}
each other within the same interval of distances $r_{break} \le r \le 
l_{max}$ where
scissions take place. While covalent bonds are thus constantly broken or
created during the simulation, we would like to emphasize that no 
formation of ring polymers is allowed and this condition has to
be observed whenever an act of polymerization takes place.

The non-bonded interaction between monomers is described by a Morse potential,
\begin{eqnarray}  \label{Morse}
U_M(r)=\exp\left[-2a\left(r-r_{min}\right)\right]-2\exp\left[-a\left(r-
r_{min}\right)\right],\quad \ for \quad 0<r-r_{\min}< \infty,
\end{eqnarray}
where $r_{\min }=0.8$, and the large value of $a=24$ makes
interactions vanish at distances larger than unity, so that an efficient
{\em link-cell} algorithm\cite{Gerroff} for short-range interactions can be
implemented. In the present study we maintain our system in the "good solvent"
regime and, therefore, only keep the repulsive branch of Eq.(\ref{Morse}), 
shifting it up the positive $y$-axis so that $U_M(r) = 0$ for $r > r_{min}$.
The radii of the beads and the interactions, Eqs.(\ref{FENE}), (\ref{Morse}%
), have been chosen such that the chains may not intersect themselves or
each other in the course of their movement within the box, so that
'excluded-volume' interactions as well as the topological connectivity of
the macromolecules are allowed for.

\begin{figure}\label{B}
\includegraphics{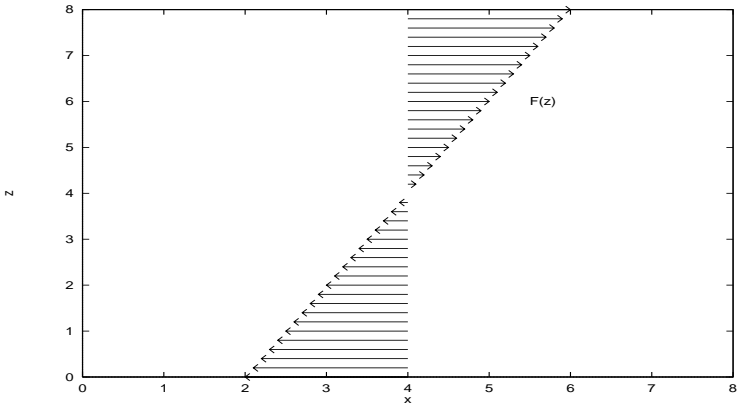}
\vskip 5.5cm
\caption{Definition sketch for external field variation, $F_x(z)$,
causing flow in $x$-direction between infinite parallel plates.}
\end{figure}

We introduce the shear rate $B$ by defining an external field $\vec{F}$ whose only 
component is directed along the $x$-axis and changes linearly along 
the $z$-dimension of the box - Fig.\ref{B}:
\begin{equation}
\label{F}
F_x(z) = B(z - Z_{max}/2), \quad \frac{dF_x(z)}{dz} = B
\end{equation}
so that the bias changes sign at the middle of the box $Z_{max}/2$.

A standard Metropolis algorithm governs monomer displacements, whereby
an attempted move of a randomly selected particle in a random direction
is taken from a uniform distribution within the interval
$-\frac{1}{2} \le \Delta x, \Delta y, \Delta z \le \frac{1}{2}$. 
The presence of impenetrable walls at $z = 0$ and $z = Z_{max}$ is
observed by rejection of all those jumps of the monomers which would
otherwise cause them to leave the box through the planes at the
bottom and the top. Thus jumps are attempted with probability 
\begin{equation}
P_{att}(\Delta x) = \left\{\begin{array}{ll}
1,\quad &\mbox{for} \quad -\frac{1}{2} \leq \Delta x \leq \frac{1}{2}\\
0,\quad &\mbox{otherwise} 
\end{array} \right.
\end{equation}
and {\em accepted} with probability, equal to
\begin{equation}
P_{acc}(\Delta x) = \left\{\begin{array}{ll}
\exp [-(E_{new} + \Delta W
- E_{old})/k_BT],\quad &\mbox{for}\ \quad
E_{new}+\Delta W>E_{old},  \label{Met} \\
1,\ &\mbox{otherwise},
\end{array} \right.
\end{equation}
where $E_{new}$ and $E_{old}$ are the energies of the new and old system
configurations, and $\Delta W$ is the  work, 
performed by the field when a monomer
jumps from position $\vec{r}_{old}$ to a new position, $\vec{r}_{new}$:
\begin{equation}\label{work}
\Delta W = \int_{\vec{r}_{old}}^{\vec{r}_{new}}\vec{F}\vec{dr}=
\int_0^{\Delta x}F(z(x))dx.
\end{equation}
Since the potential of the field $\vec{F}$, Eq.(\ref{F}), is not scalar, 
$\Delta W$
depends on the path followed by the monomer during a jump $\vec{r}_{old}
\rightarrow \vec{r}_{new}$.  During a jump this path $z(x)$ is a straight line, 
$z(x) = \frac{\Delta z}{\Delta x}x+z_{old}$, so that from Eq.(\ref{F}) 
one obtains for the work
\begin{equation}\label{aver}
\Delta W = \int_0^{\Delta x} F_x.dx = B\frac{z_{new}+z_{old}}{2}\Delta x =
\bar{F}\Delta x, 
\end{equation} 
where $\bar{F}$ denotes the {\em average} from the values of the field in positions 
$\vec{r}_{old}$ and $\vec{r}_{new}$. With $\bar{F}$, as defined in Eq.\ref{aver}, one 
satisfies the condition of {\em microscopic reversibility} with respect to 
the movements of the particles. Note that it is the microscopic reversibility
which requires the use of $\bar{F}$, rather than $F_{old}$, for instance, in
the determination of $\Delta W$. 
This becomes immediately obvious by considering the displacement of a single 
non-interacting particle in the external field: for a self-consistent algorithm, 
the update rules for the forward and reverse moves have to be identical
(time reversible).

The field, $\vec{F}$, is thus introduced in the system as an additional
term in the Boltzmann probability in Eq.(\ref{Met}) whereby the probability
for jumps along or against the field becomes then strongly biased.
This term makes the energy
in the Boltzmann factor in Eq.(\ref{Met}) a monotonously decreasing function
of $x$. The Metropolis algorithm tends to find the minimum of this energy thus
continuously driving the system to an unreached minimum which gives a
continuous flow in $x$ direction. One can readily estimate the average
jump distance, $\delta x$, if $\Delta W$ is assumed to be larger than the 
microscopic interactions $U_{FENE}$ and $U_M$. Setting $E_{new} \approx
E_{old} = 0$ for simplicity in Eq.(\ref{Met}), with $\bar{F}\approx F$ one has
\begin{equation}
\label{xmean}
\delta x = \frac{\int xP_{acc}(x)dx}{\int P_{acc}(x)dx} = 
\frac{\int_{0}^{\frac12}x\exp(-Fx)dx + \int_{-\frac12}^{0}xdx}
{\int_{0}^{\frac12}\exp(-Fx)dx + \int_{-\frac12}^{0}dx} =
\frac{-\frac18 + [1-(1+\frac{F}{2})\exp(-\frac{F}{2})]/F^2}
{\frac12 + [1-\exp(-\frac{F}{2})]/F}. 
\end{equation}
which yields $\delta x = 0$ for $F = 0$ and $\delta x \approx -\frac14$ 
for sufficiently strong fields $F$,
(the sign of $\delta x$ depending on which half of the box is considered).
Thus $\delta x$ remains bound from above, no matter how strong the
applied bias is chosen\footnote{Of course, since the microscopic interactions 
are always present, this result remains a rough estimate only whose validity 
can be checked by the simulation.}. 
 
In order to provide for the reversibility of scission and recombination events, 
a Monte Carlo time step (MCS) has been performed after $N$ particles of the 
system are randomly chosen and attempted to move at random whereby existing 
bonds are kept as they are. After that $N$ monomers are again picked at random, 
i.e. one of the two bonds of each is randomly chosen and, depending on whether 
this bond exists and points to an existing neighbor along the same chain, 
or it doesn't, the bond is attempted to break or to create.

Thus, within an elementary time step (MCS) $N$ random jumps and as many attempts
of bond scission/recombination are carried out, each subject to the Boltzmann
probability that the respective attempt is successful. 
It is clear that in a system of EP where scission and recombination of bonds 
are constantly taking place the particular scheme of bookkeeping is no
trivial matter. Since the identity of a particular chain, or monomer affiliation,
is in principle preserved for no more than one MCS, the data structure of the 
chains can only be based on the individual monomers (or, rather, links) as
suggested recently\cite{WMC2}. Thus each monomer has two links.
The links associated with a given monomer  are pointers which may either point
In the latter case a link then represents an unsaturated dangling bond.
Thus a large number of particles may be simulated at very modest
operational memory. Results in the present study involve systems of up to
$N = 65536$. The simulational box (slit) sizes are typically $16 \times 16
\times D$ where $D$ is the width of the slit in $z$-dimension. The total
density of the monomers $c$ is then defined as the number of monomers per
unit volume.
 
During the simulation the whole system is periodically examined,
the number of chains with chain length $l$, square end-to-end distance, 
$R_e^2$, gyration radius, $R_g^2$, center-of-mass coordinates, 
displacements, etc., are counted and stored. Because of the semi-periodic 
(in $x$- and $y$-direction) boundary conditions the interactions between 
monomers follow the minimum image convention. The computation of the 
conformational properties of the chains as $R_e^2$, for instance, then implies
a restoration of the {\em absolute} monomer coordinates from the periodic ones
for each repeating unit of the chain. 

Technical details of this new
algorithm will be presented elsewhere, here we will note only that the high 
efficiency in code performance is achieved by extensive implementation of 
integer arithmetic in this off-lattice model based largely on binary 
operations with variables. 
Thus, for example,  the most heavily 
involved (modulo) operations which provide periodicity of coordinates and the
minimum image computation for distances turn out to be redundant.

%% file: bias.tex
\section{Simulational Results}
\label{shear}
\subsection{Velocity Profiles}
\begin{figure}
\includegraphics{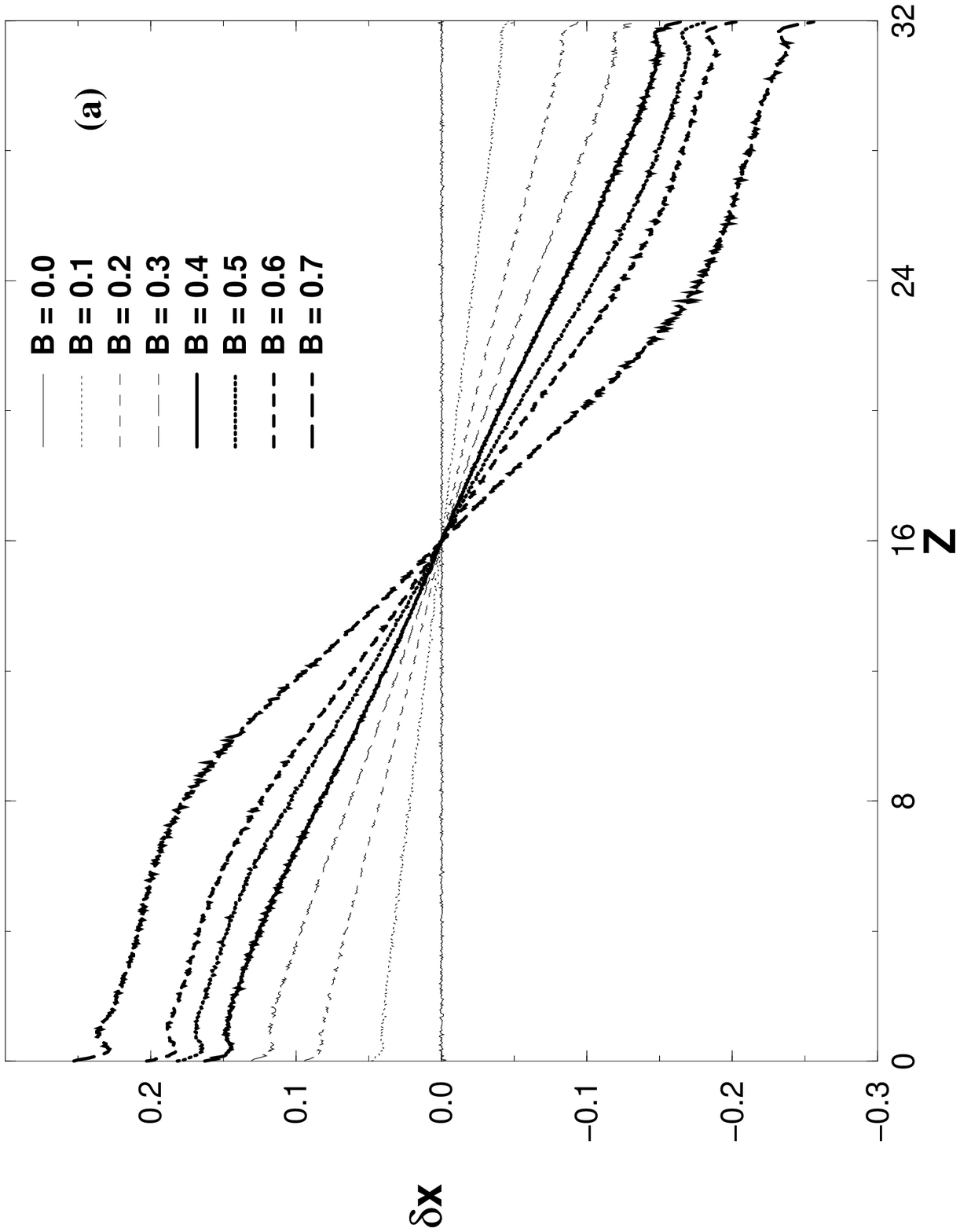}
\includegraphics{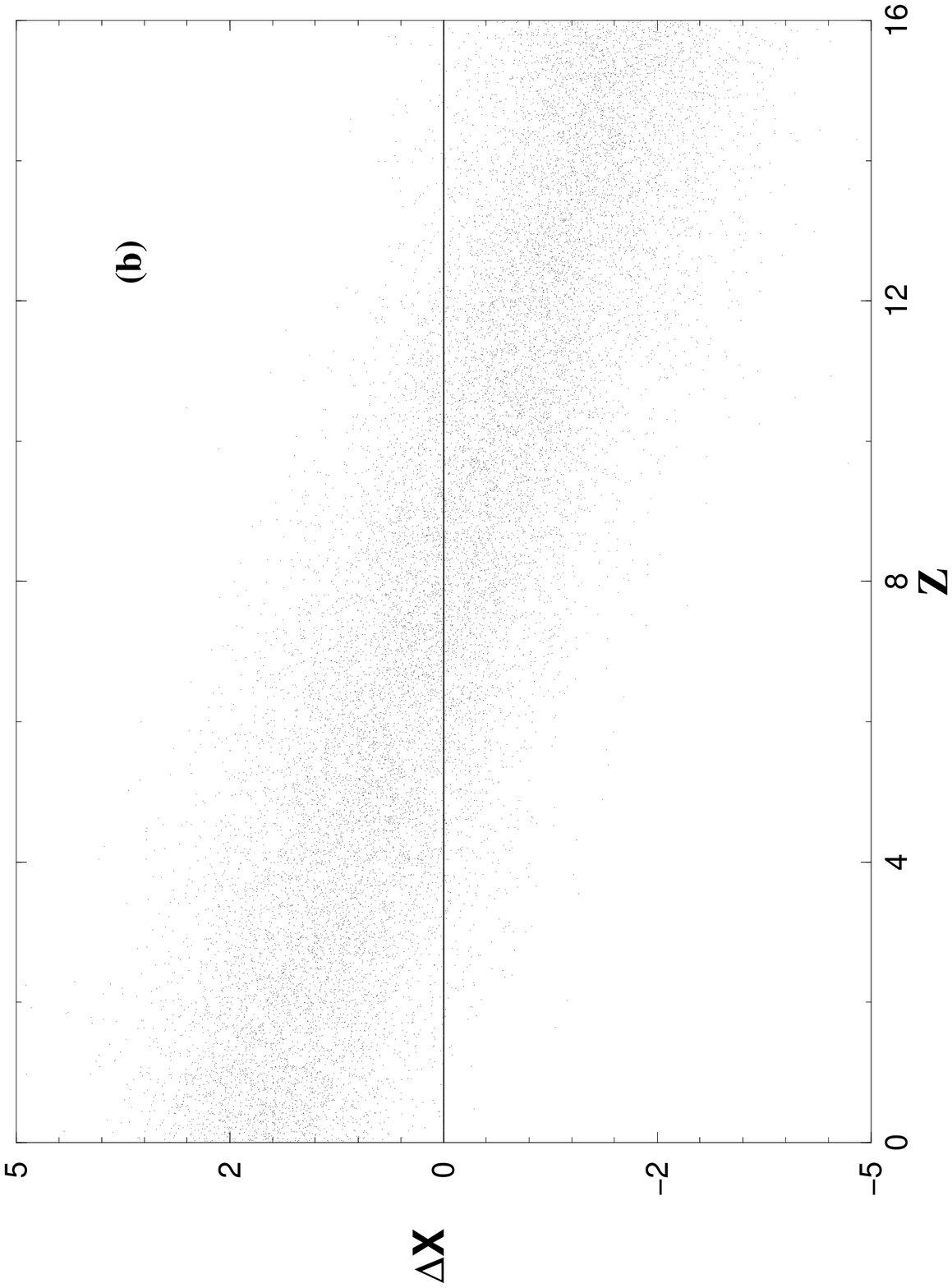}
\vskip 5.5cm
\caption{\label{jumps}(a) 
Variation of the average distance of {\em accepted} jumps
in $x$-direction vs $z$ coordinate with the external field amplitude $B$ 
in a box of size $Z_{max} = 32$. (b) Variation of the average velocity
(distance in $x$-direction traveled by a monomer after $1024$ MCS)
at density $c=1.0, Z_{max}=16$, and $B=0.1$.}
\end{figure}

As mentioned in Section \ref{model}, we create a shear flow in our system 
by applying an external field with constant gradient along the $z$-axis of
the box so that the flow is oriented along the $x$-direction - cf. 
Eq.(\ref{Met}), parallel to the hard walls at $z = 0$ and $z = D$. 
The lower half of the box would then flow in positive, the upper one - in
negative $x$-direction. It is expected that in the immediate vicinity of
the walls the flow might be somewhat distorted due to walls impenetrability. 
Below, in Fig.(\ref{jumps})a we plot the mean jump distance per MCS, 
$\delta x$, measured along the $x$-axis for different values of $B$. 
The $z$-coordinates of these successful jumps are taken from the respective 
$z$-coordinate of the monomers. 

Evidently, in a wide channel with $D = 32$ only for sufficiently weak field $B 
\le 0.3$ the average jump distance grows linearly with respect to the half-width
of the box (for $B = 0$ it is zero). 
For $B > 0.3$ distortions in the $\delta x$ profile set in because the maximal 
jump distance is limited to $0.5$, as mentioned in Section \ref{model}.
For a more narrow slit of width $D=16$ the region of 
linear response would then extend to higher values of $F\le 0.7$. Therefore
most of the simulational results in what follows are derived for $D=16$.
Fig. \ref{jumps}b then demonstrates that the velocity changes linearly 
across the slit for sufficiently small values of the field $B$.

In the broad channel, $D = 32$, at $z = 0, D$ for $B = 0.5$ one gets $F 
= 8$ from Eq.(\ref{F}) so that the average jump distance there according to
Eq.(\ref{xmean}) should be $\delta x \approx \pm 0.178$. The value of 
$\delta x$ at the borders of the box,
as seen from Fig.(\ref{jumps})a, confirms this estimate demonstrating
that the role of the microscopic interactions $U_M$ and $U_{FENE}$ is small.

The presence of the walls is felt in their immediate vicinity and some local
distortion of the displacement profile appears increasingly pronounced with
growing $B$ although it remains spatially contained in a layer of thickness 
roughly equal to monomers diameter. It is interesting to note that this small
increase of $\delta x$ (and, therefore, of velocity) immediately at the 
walls resembles the so called "slip effect"\cite{Biller,Goh} in simple
shear flow of dilute polymer solutions in a narrow channel. This slip effect
can be explained intuitively by the fact that the polymer molecules near the 
wall align themselves more strongly with the flow than those away from the 
wall, and are thus transporting less flow-wise momentum across the flow
than would otherwise be the case. Indeed, in our Monte Carlo model the 
attempted jumps which would otherwise bring the monomers through and beyond 
the walls of the slit are always rejected. Since the molecules cannot
penetrate the wall, their concentration is reduced at the wall, so that their 
contribution to the viscosity is further diminished. Using a model of dumbbells 
in parallel wall shear flow, one can calculate\cite{Aubert,Biller,Goh} both the
nonlinear velocity profile of the suspended solutions as well as the 
center of mass concentration profile between the two walls - we shall 
see in the next section that the latter is qualitatively reproduced by
our simulational results. 

\subsection{Effect of Shear Rate on Average Chain Length and Molecular
Weight Distribution}
\begin{figure}
\includegraphics{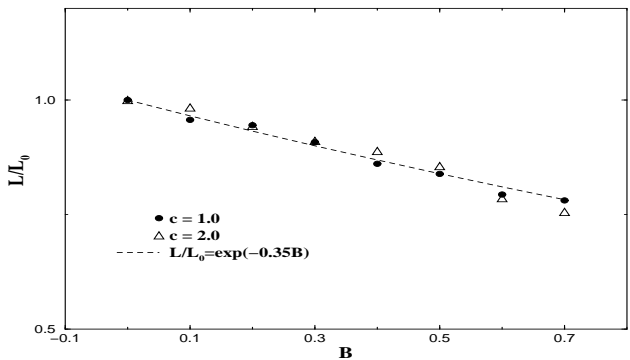}
\vskip 5.5cm
\caption{\label{lvsb} Relative decrease of mean chain length $L$ in lamellar
flow field versus $B$ at $J/kT = 7$, densities $c = 1.0$ and $c = 2.0$ and 
$D=16$.}
\end{figure}

\begin{figure}
\includegraphics{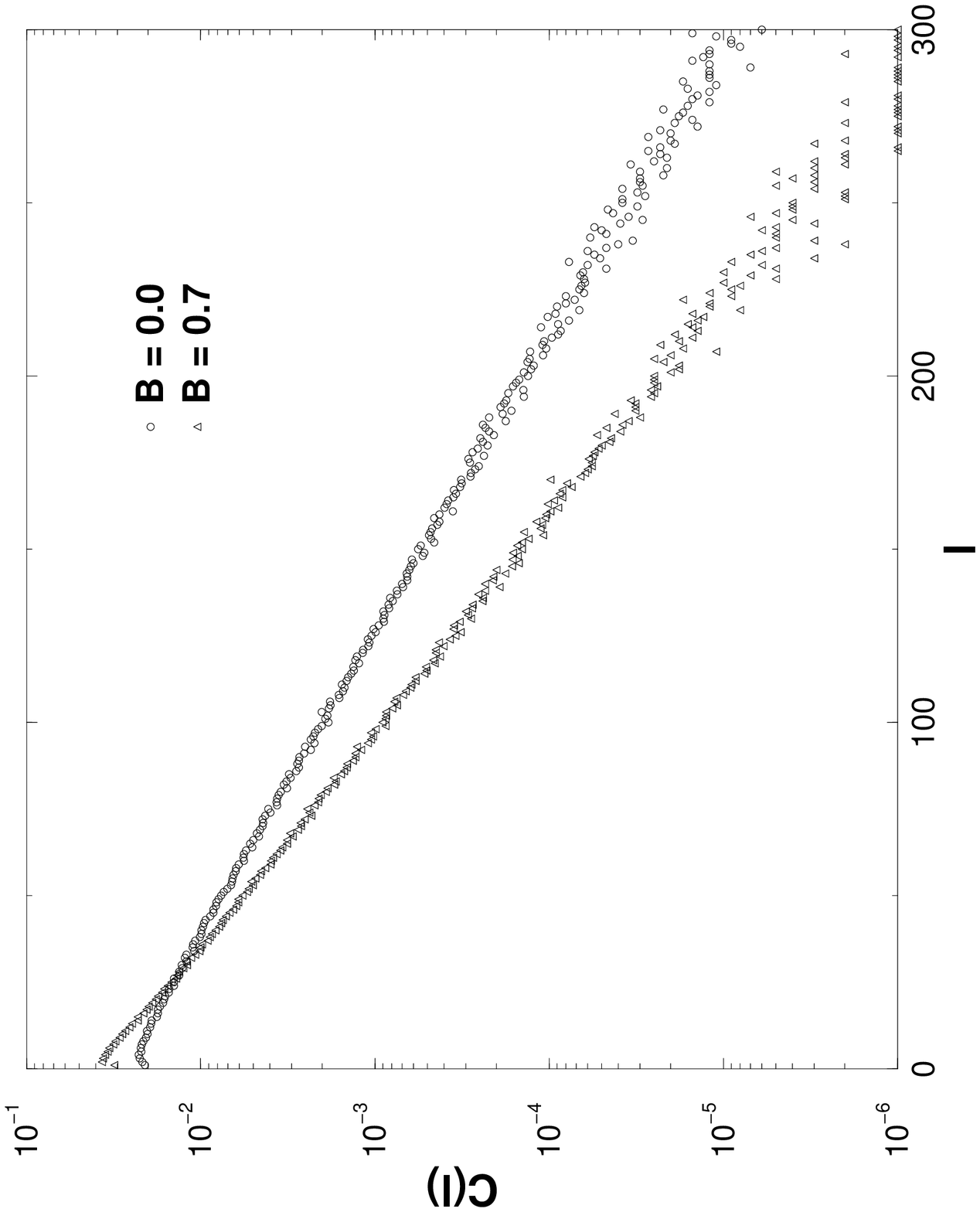}
\includegraphics{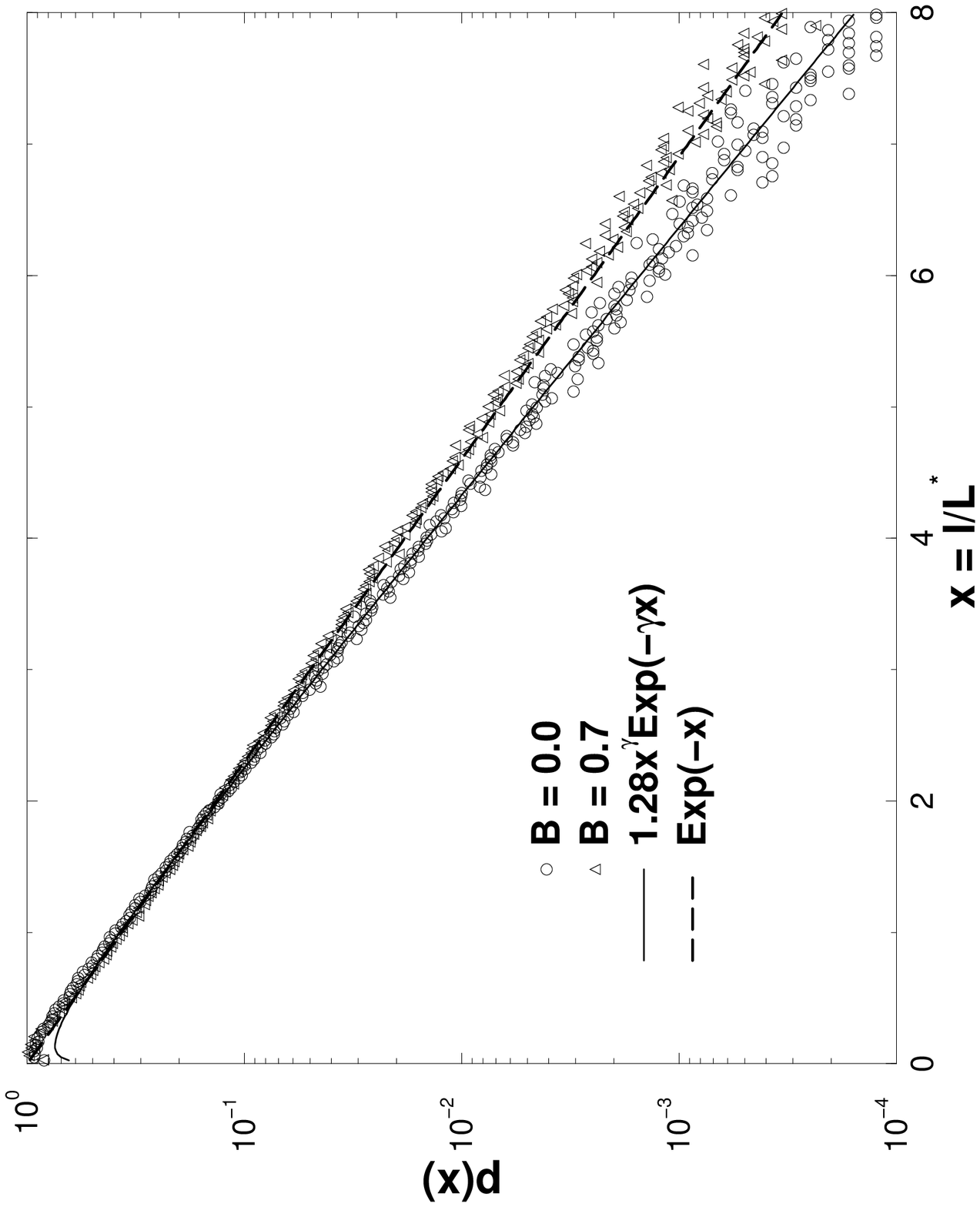}
\vskip 5.5cm
\caption{\label{mwd} (a) Molecular weight distribution in a system of EP 
at rest, $B = 0$, and in a flow, $B = 0.7$, at $J/kT = 7$ and monomer density 
$c = 0.5$. (b)The same as in (a) in dimensionless units $x=l/L$ fitted with
Eq.(\protect\ref{MWD}) with $\gamma = 1.16$. }
\end{figure}

We find that the average chain length of the EP solution, $L$, decreases 
steadily with growing $B$, which is in agreement with an earlier MD 
study\cite{K} - Fig. \ref{lvsb}:
The mean chain length $L$ at this highest shear rate is about $70 \%$ of 
its value for a system at rest. Here we should like to point out the 
existence of considerable fluctuations in the derived values of $L$ for
$F \neq 0$ - the statistical error has been reduced at the expense of 
considerable computational effort.
Note that the reduced $L/L_0$ mean
chain length dependence on shear rate $B$ ($L_0$ is the mean chain length 
of the solution at $B = 0$) appears to decrease nearly linearly with $B$:
$L/L_0 = 1 - 0.35 B$ which simply follows (cf. Eq.(\ref{L}) below) from the
exponential dependence of $L$ on interaction $J$: $L/L_0 = 
\exp[-(J-bB)/2]/\exp(-J/2) = \exp(bB/2) \approx 1 - bB/2$ with the constant
$b = 0.7$ measuring the effective decrease of bond energy $J$ due to shear $B$.
Another interesting observation is that the rate of decline is apparently 
independent of density $c$, at least for the small values of $F$ considered
in the present work. 
The form for the MWD $C(l)$, 
that is, the concentrations of chains of contour length $l$, appears to change
qualitatively for equilibrium polymers in dilute solutions. 
This change is in line with the predictions of the recent scaling theory
of EP\cite{WMC2} where we demonstrated that the purely exponential form of the 
MWD, $P(x) \propto \exp(x)$, corresponding to concentration/chain length
regimes in which density correlations are suppressed (typically beyond the 
semi-dilute
threshold), is
replaced by a 'rounded' Schwartz power-exponential distribution:
\begin{equation}
P(x) dx =
\left\{ \begin{array}{ll}
\exp(-x) dx & \mbox{($L \gg L^*$)} \\
\frac{\gamma^{\gamma}}{\Gamma(\gamma)} x^{\gamma-1} \exp(-\gamma x) dx
        &\mbox{($L \ll L^*$)}
\end{array}
\right.
\label{MWD}
\end{equation}
when correlations are important (typically dilute concentrations).
In Eq.(\ref{MWD}) the reduced chain length, $x = l/L$, is taken as ratio 
of the particular chain length $l$ to the mean
chain length $L$, $\gamma$ is the critical exponent of the $n\rightarrow 0$
vector model, (in 3D $\gamma \approx 1.165$ while its mean field value is $
\gamma_{MFA} = 1$), and $L^*$ marks the average
chain length at the crossover from dilute to semi-dilute concentration, 
$(c \rightarrow c^*)$, of EP solutions. 
The mean chain length $L$ was predicted
and confirmed to vary with dimensionless bond energy $J/k_BT$ as
\begin{equation}
L  \propto c^{\alpha} \exp(\delta J/k_BT)
\label{L}
\end{equation}
with exponents
$\alpha_d=\delta_d=1/(1+\gamma)\approx 0.46$ in the dilute and
$\alpha_s=1/2 (1+(\gamma-1)/(\nu d-1) \approx 0.6$, $\delta_d =1/2$ in the
semi-dilute regime.
In Fig. \ref{mwd} we 
plot $C(l)$ for a system at rest ($c = 0.5,\ B = 0$) and at maximum shear 
rate ($B = 0.7$) to demonstrate that the form of MWD changes qualitatively 
when shear is imposed.
Thus it is evident from Fig. \ref{mwd} that in the presence of shear the 
correlations in polymer concentration
in our dilute system of EP are effectively suppressed and the Molecular Weight
Distribution is very well reproduced by the simple exponential function expected
if a MFA description of the system holds. This finding can be understood
if one recalls that the imposition of an external field with a shear rate
$B$ has a twofold effect on the polymers: (i) it effectively reduces the bond strength,
$J \rightarrow J - b B$, which makes the polymers shorter, and (ii) the
shape of the polymer coils is changed towards more rod-like shape with the longest
axis oriented along the field. Meanwhile it is well known that a system of
rods exhibits a Mean-Field-like behavior\cite{CC} which is here manifested
by the change in the MWD - Fig.\ref{mwd}.

\subsection{Effect of Shear Rate on Chain Conformations}

\begin{figure}
\includegraphics{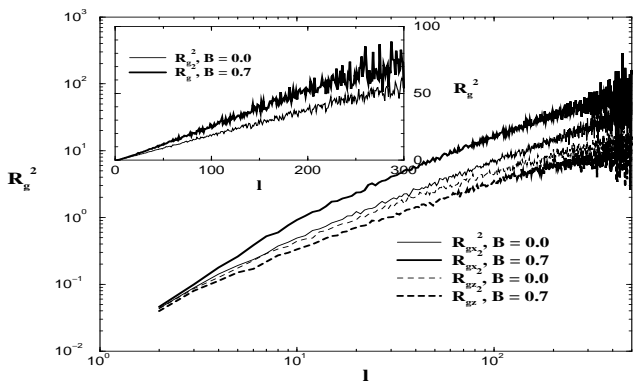}
\vskip 5.5cm
\caption{\label{frg2} Variation of coil size along the field, $R_{gx}^2$, and
perpendicular to field, $R_{gz}^2$, with chain length $l$ and two values of
$B = 0.0,\; \mbox{and}\; 0.7$. The inset shows the {\em total} $R_g^2$ in normal
coordinates. Here $J/kT = 7$ and the density $c = 0.5$ in a narrow slit $D = 16$.
The average coil size measured  at zero shear rate is $R_e^2 = 50.03$ and 
$R_g^2 = 8.325$.}
\end{figure}

As the flow becomes faster, the individual shape of the chain coils
change too. For weak shear rates $B$, the relative distortions (as measured
for instance by flow birefringence\cite{J}) are essentially proportional
to $\tau B$ where $\tau$ is the largest relaxation time of the unperturbed
molecule\cite{Zimm}. Brownian dynamics
studies\cite{Diaz} for Hookean dumbbells in a steady shear flow confirm this
relative increase of the end-to-end distance with the shear rate both with
and without hydrodynamic interactions included. A confirmation of these early
predictions follows from our simulational results too - Fig.\ref{frg2}. 
\newpage
In Fig.\ref{frg2} we plot the difference in gyration radii, typical for a 
large section of the length distribution and averaged over all EP, in two 
characteristic cases - system at rest and in a flow, 
indicating that chain coils in a flow become more extended along the field
and compressed parallel to it. The result is a total increase of $R_g^2$ as
the whole system starts drifting along field's direction. It is also
evident from Fig. \ref{frg2} that this asymmetry in the components of 
$R_g^2$ becomes progressively more pronounced as $l$ gets larger. The fact that
$R_{gz}^2$ is somewhat smaller than $R_{gx}^2$ even at rest is due to the presence of 
hard walls at $z = 0, Z_{max}$ which slightly deforms the coils in $z$ -
direction.
We have not given here a plot of the $z$-dependence of $R_g^2$ because of
the surface segregation of chain lengths, induced by the parallel plates.
This segregation populates the vicinity of the walls with single monomers and very short
species. In contrast, the longer chains reside at least a distance $R_g$ always
from the walls (see next Section). Such a distribution of centers of mass
with respect to chain length takes place in EP even at rest (and is further 
enhanced by the flow) making the MWD a $z$-dependent quantity and thus
interfering with the pure effect of coil stretching under flow.

\subsection{Density Profiles}

\begin{figure}
\includegraphics{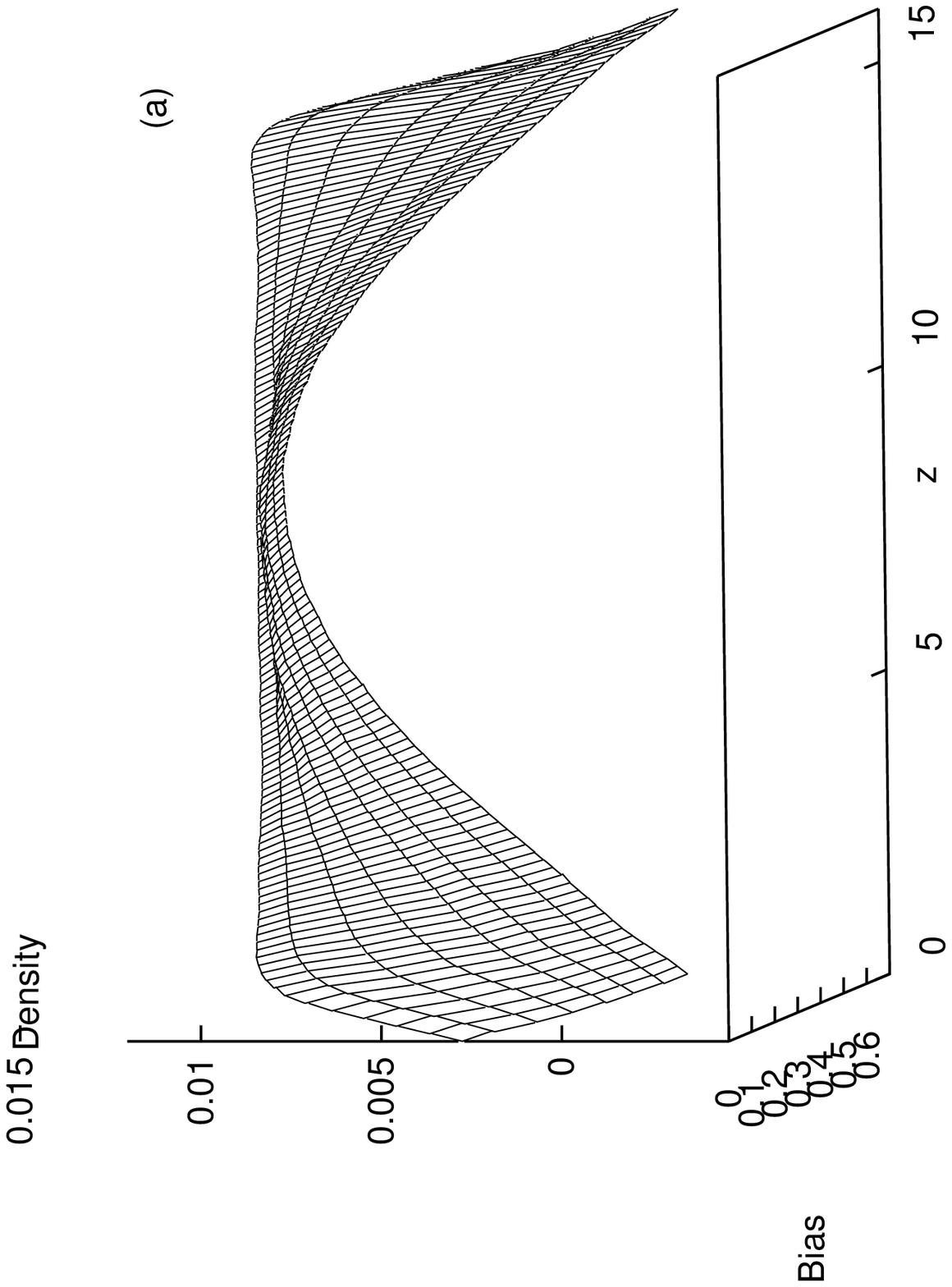}
\includegraphics{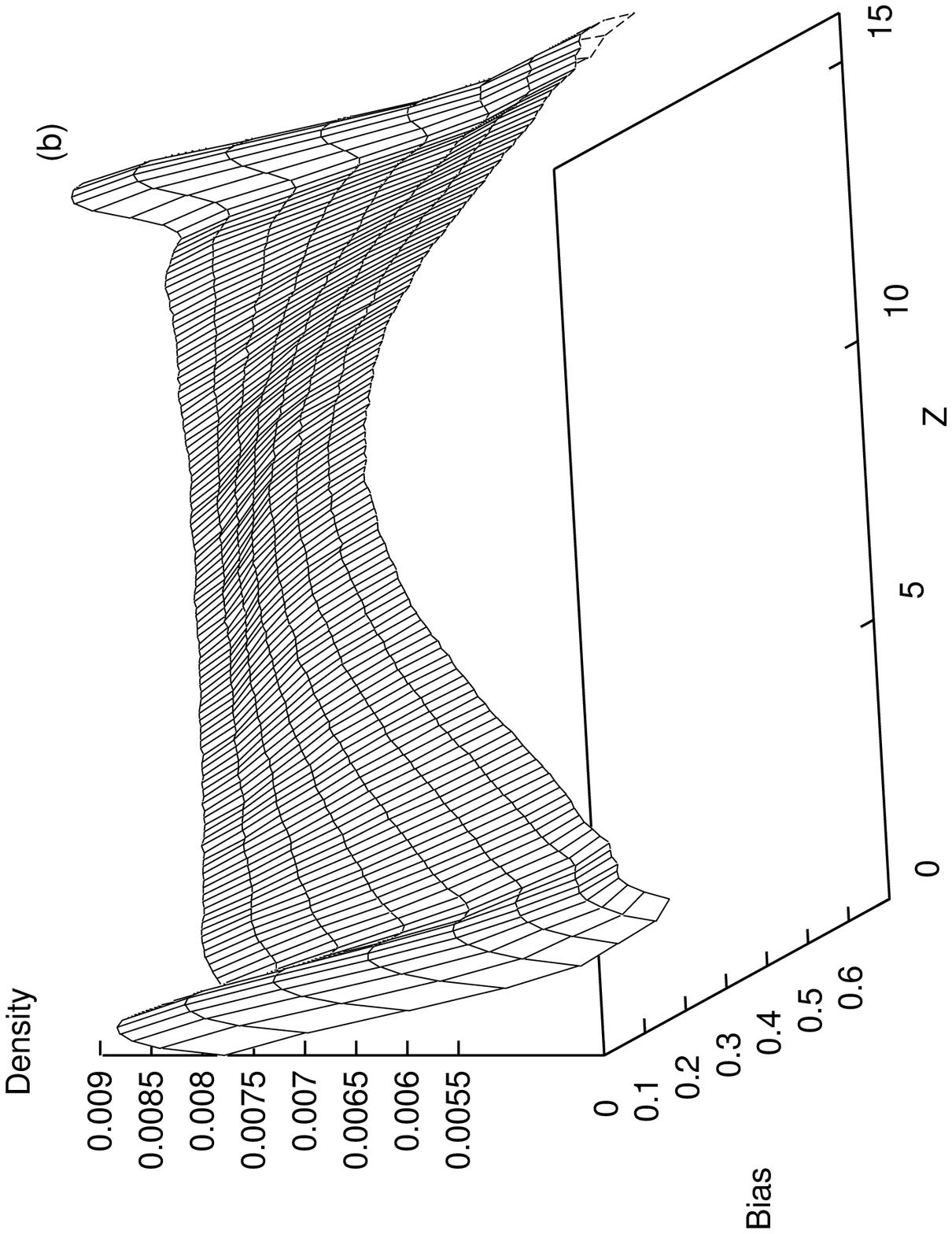}
\vskip 5.5cm
\includegraphics{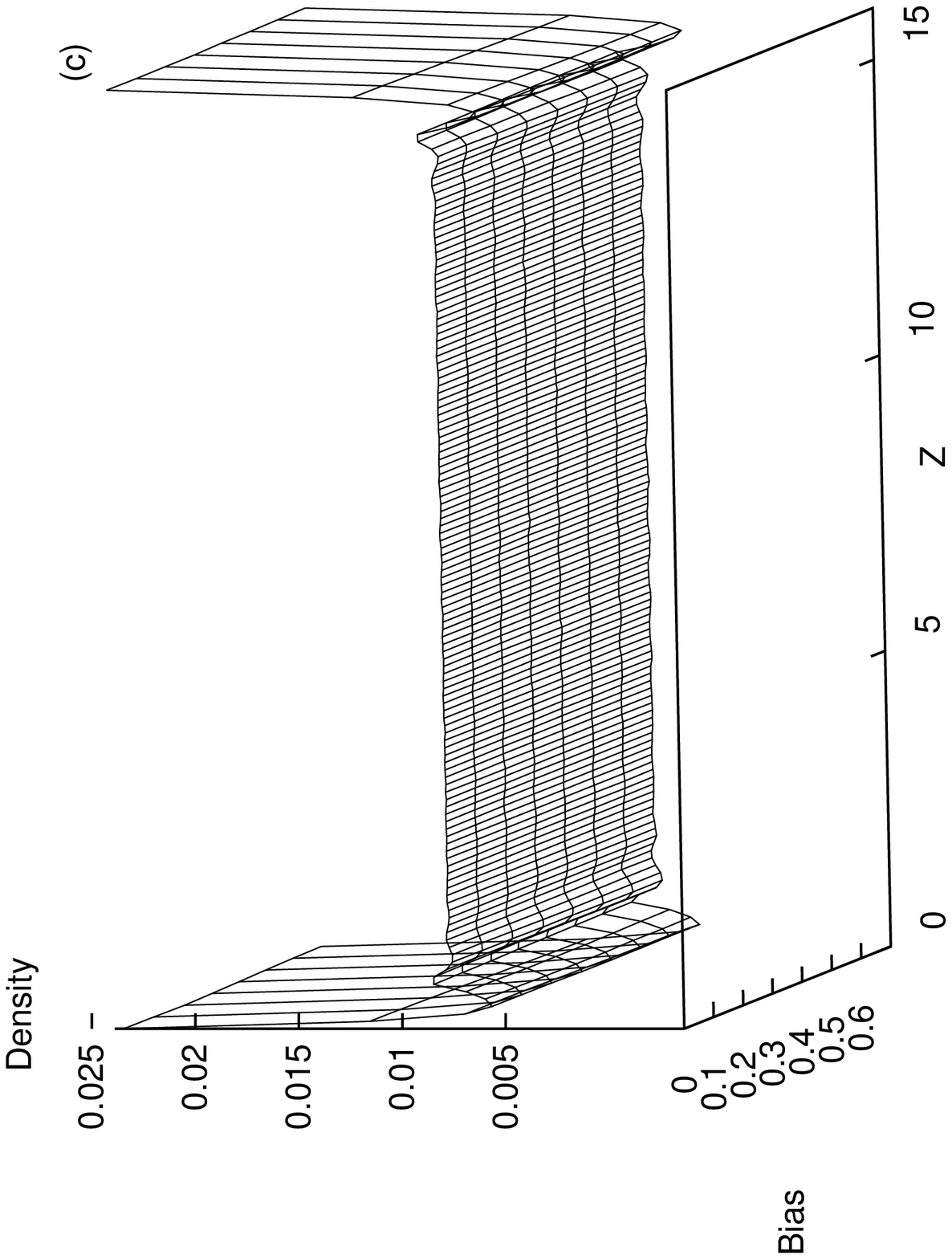}
\includegraphics{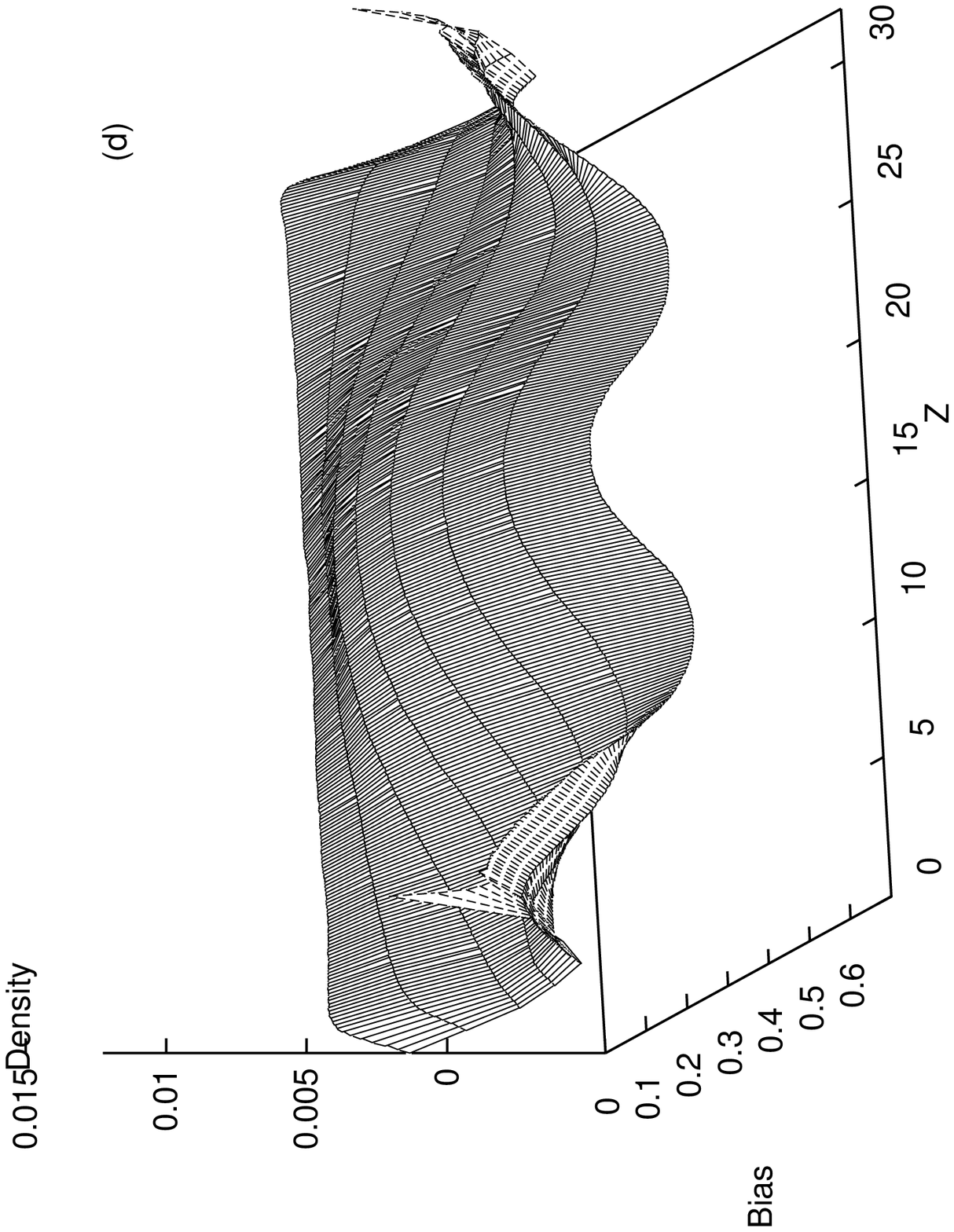}
\vskip 5.5cm
\caption{\label{dens} Distribution of total monomer density between the 
hard walls in a slit of width $Z_{max} = 16$ at varying shear rate (bias) and
$J/kT=7$: 
(a) $c = 0.5$; (b) $c = 1.0$; 
(c) $c = 2.0$; (d) a broad slit with $Z_{max} = 32$ and $c = 0.5$. 
}
\end{figure}

The overall transformations which the system undergoes with increasing
shear rate, however, become much more explicit if density and diffusion
profiles are sampled as function of $z$. This is shown in Fig.\ref{dens}
where the density is normalized to unity ($\int_0^D c(z) dz = 1$).
It is evident from Fig.\ref{dens}a that in the absence of bias when the 
system is at rest the total monomer density is uniformly distributed 
across the box with a typical depletion immediately at the walls (at
low concentration of the solution). The walls are avoided by the
longer chains because of entropic reasons. When the system starts to flow 
a redistribution of density sets in with increasing bias $B$ whereby for 
the highest shear rate one observes a density maximum centered at the middle 
where the flow velocity is nearly zero. Qualitatively this density profile
appears to be similar to analytic and simulational results\cite{Goh} for the
center of mass concentration profile between upper and lower walls,
obtained earlier for a single dumbbell in a slit\footnote{In the much
simpler model\cite{Goh}, however, the density profile is independent of
the shear rate $B$.}.  

As the concentration is further increased, one observes the onset of typical
oscillations in density profiles in the vicinity of the walls, Fig.\ref{dens}b,
c. Such oscillations are typical for polymer solutions confined between 
flat plates and have been comprehensively studied for conventional
polymers by Monte Carlo simulations before\cite{Yet,PMB}. The observed transitions 
in the monomer density immediately at the walls from a deficit (depletion
layer) at low concentration up to an excess (for melts) are governed by a
competition between entropic and packing effects. Because of the 
resultant decrease of configurational entropy it becomes unfavorable for 
polymers to be near the walls.
Chains near the walls, on the other hand, suffer collisions with the chains 
away from the walls, and tend to move closer to
the walls. At low density the entropic effect dominates, 
while at high densities packing effects prevail. This is clearly seen
in Fig.\ref{dens}a,b,c for zero shear. In dilute solutions, as seen from
Fig.\ref{dens}a, the increase of shear rate leads to effective broadening
of the depletion layers, adjacent to the walls, which is in agreement with 
EWIF (evanescent wave-induced fluorescence) experimental 
observations\cite{Aussere2} of GM, but at variance with an earlier
computer simulation\cite{Duering}.

This density variation across the slit, caused by the shear rate, appears 
to depend essentially on the overall concentration of the system.
At larger shear a density maximum still forms for the slowest layer of
flow in the box middle at concentration $c = 1.0$ whereas for very dense 
systems,
Fig.\ref{dens}c, this density redistribution with shear is suppressed. 
One may thus conclude that effects of shear on the density profiles in
a slit depend essentially on the free volume in the system which is
available for rearrangement of the polymer chains. In the
broader slit, Fig.\ref{dens}d, where the shear rate gradually diminishes 
in the vicinity of the wall (adjacent layers flow with nearly equal velocity), 
the density profile gets more complex with two
local minima and a sharp increase at the walls. This complex picture indicates
that monomer density is generally increased in locations of zero flow or 
steady flow with vanishing shear.
Clearly, the changes in these profiles with $B$ when shear sets on reflect some 
complex reorganization in the polydisperse system of EP whereby the "rapids" 
of the flow may act differently on chains of different length. 

\begin{figure}
\includegraphics{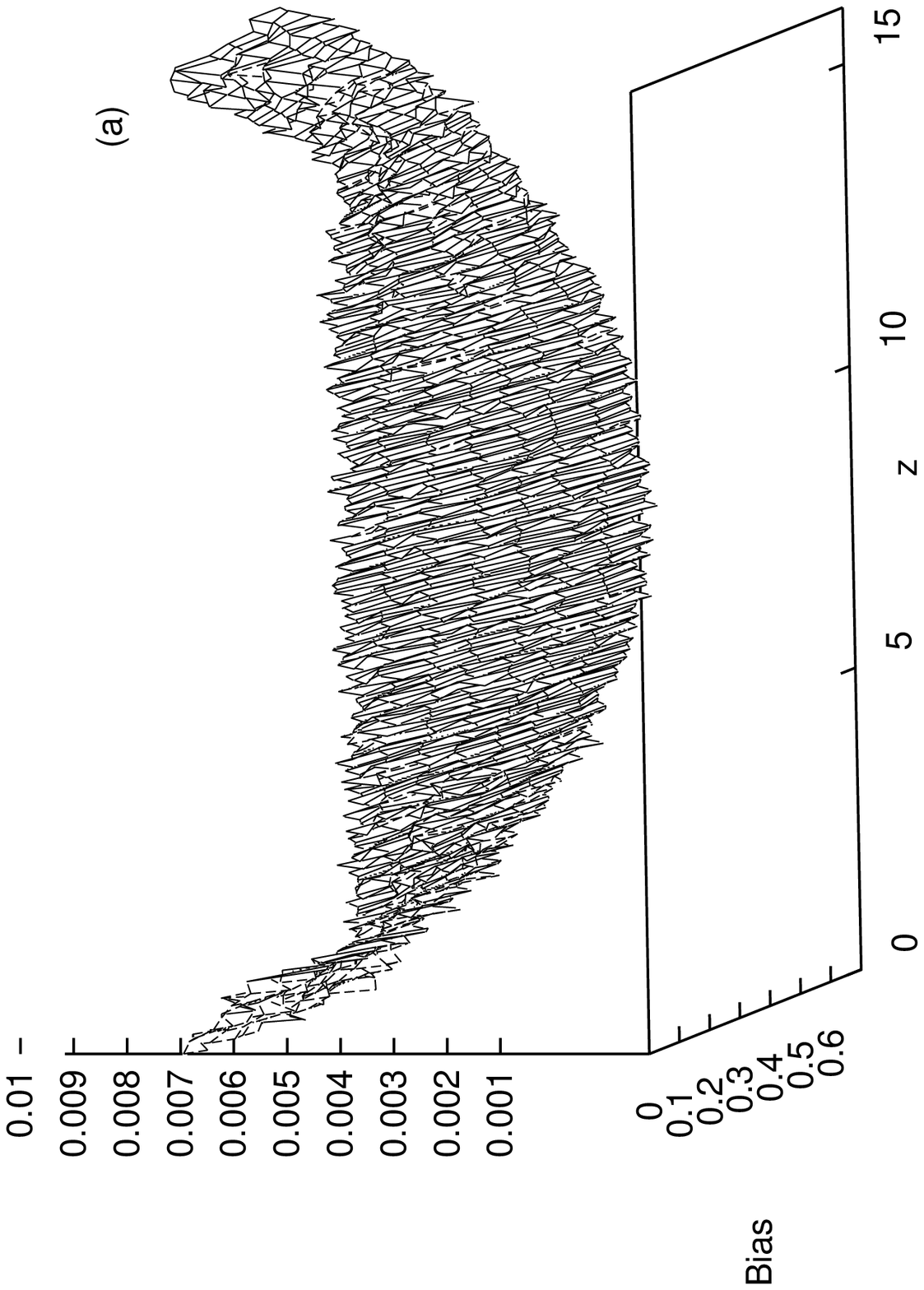}
\includegraphics{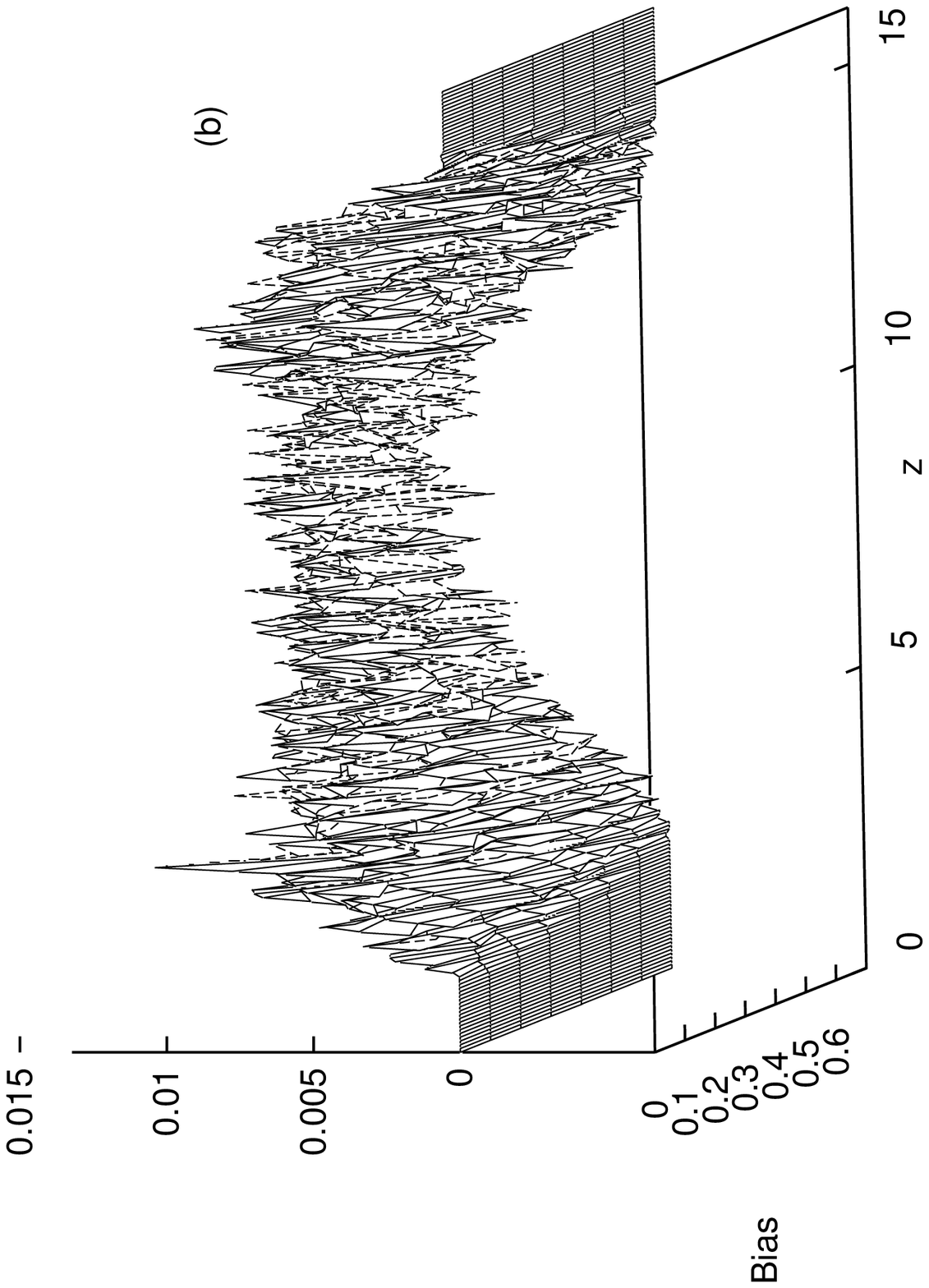}
\vskip 5.5cm
\caption{\label{zcm} (a) Center of mass distribution of single monomers in
a slit with $D = 16$, $J/kT=7$ and $c = 0.5$ at various values of the shear 
rate (bias). (b) Center of mass distribution
of chains with $l = 70$ under the same conditions as in (a).}
\end{figure}

\begin{figure}
\includegraphics{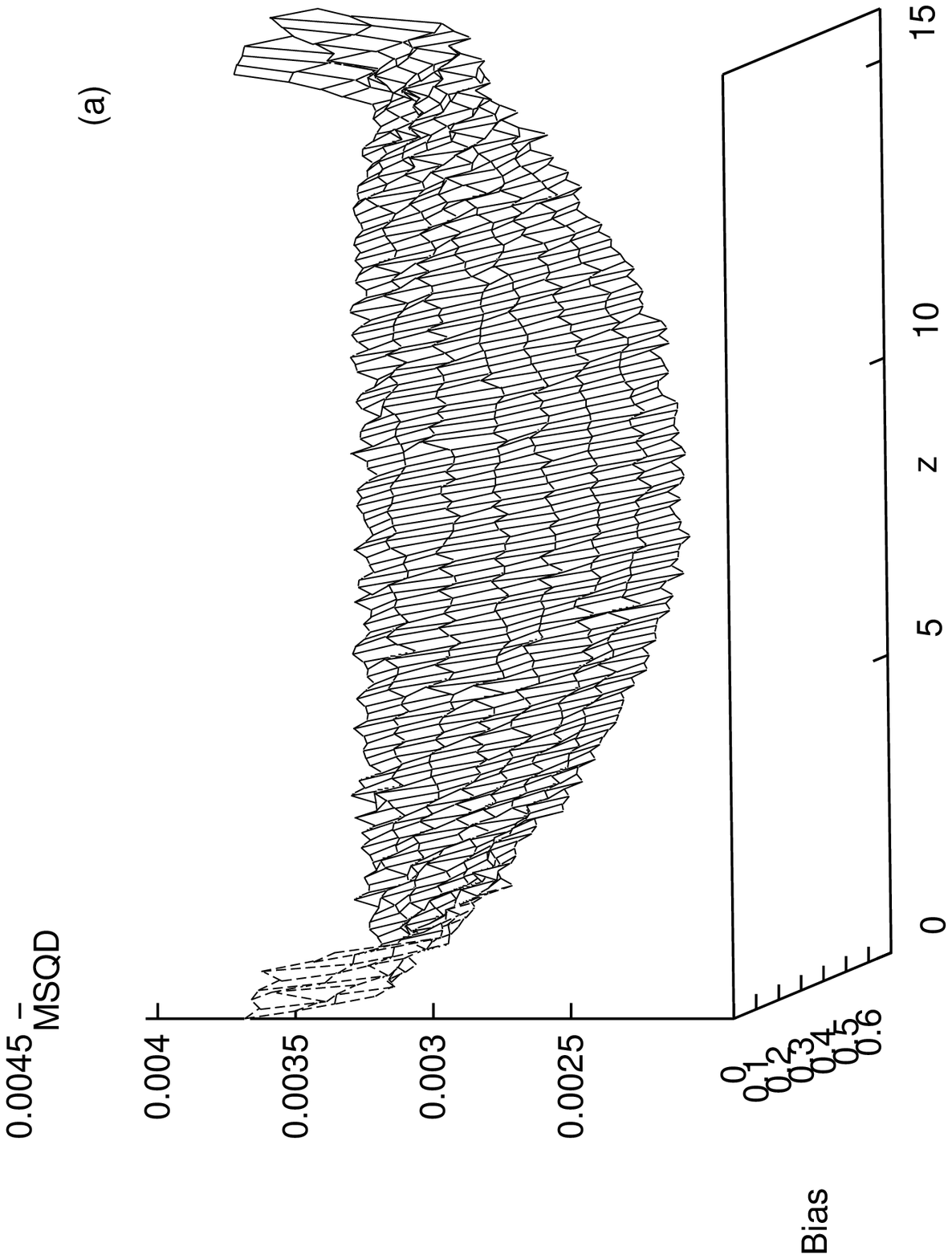}
\includegraphics{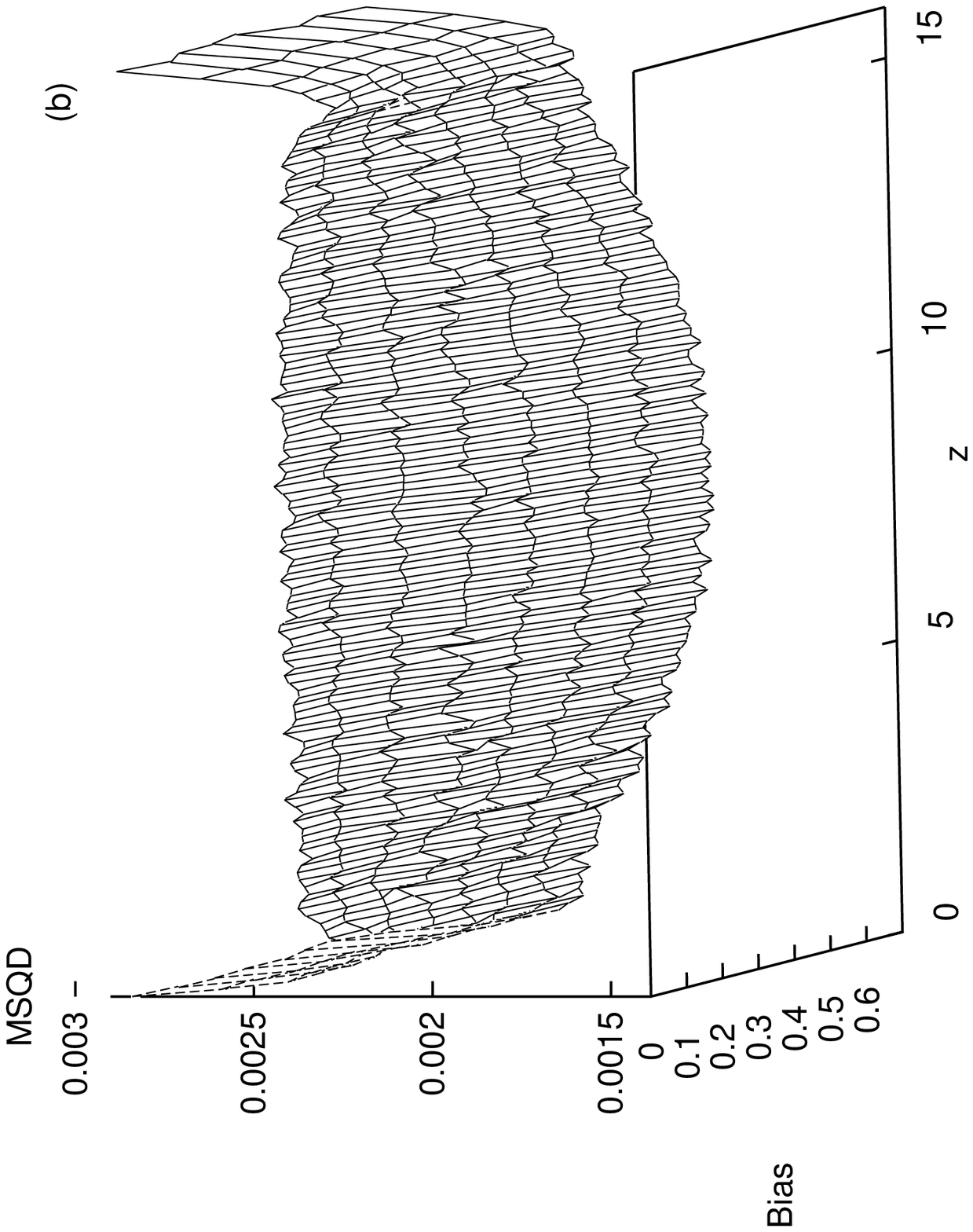}
\vskip 5.5cm
\caption{\label{dis} Distribution of MSQD during $1$ MCS in a box with 
$Z_{max} = 16$ for different shear rates (bias) at $J/kT=7$: (a) for $c = 0.5$; 
(b) for $c = 1.0$. }
\end{figure}

Additionally,
even at rest, the system segregates in the vicinity of the walls for 
entropic reasons\cite{ML} into layers occupied predominantly by chains of
decreasing contour length as one gets closer to the wall. These effects are 
indeed seen in Fig.\ref{zcm}, where the average positions of the centers of 
mass of single monomers and of chains of length $l = 70$ are shown
at various strengths of the bias field. The single monomers evidently tend to
occupy the immediate vicinity of the walls and this tendency is enhanced
as the shear increases. The long chains, on the contrary, keep at 
distance $\approx R_g$ from the walls while the system is at rest. For 
growing $B$ their residence is further narrowed around the "slow" region 
in the center of the box. In view of Fig.\ref{dens}a one may conclude
that the deficit of single monomers from this region is more than compensated 
by accumulation of longer chains. In the wide slit $(D = 32)$, only a
fraction of the long chains still remains in the middle whereas two new 
maxima at the walls appear. Evidently, this happens in those regions
where the shear rate for $B > 0.4$ (cf. Fig.\ref{jumps}) nearly vanishes. 
One might expect that other properties of the system, related to density, 
will also be affected
by the shear, as for instance, the local diffusion coefficient. In 
Fig.\ref{dis} we plot a histogram of mean square displacements (MSQD), performed
by all those particles which remain in the same $z$-layer within a MCS. 
The distribution of MSQD, Fig. \ref{dis}b, develops from being nearly constant
(with two small wings at the depletion zones) for zero bias to a well
defined broad minimum in the middle of the box as $B \rightarrow 0.7$ 
whereby as a whole it also decreases. Evidently, the diffusion profile 
across the channel reflects simply the variations of the density distribution.

\subsection{Nematic Ordering of Short Chains}
 
\begin{figure}
\includegraphics{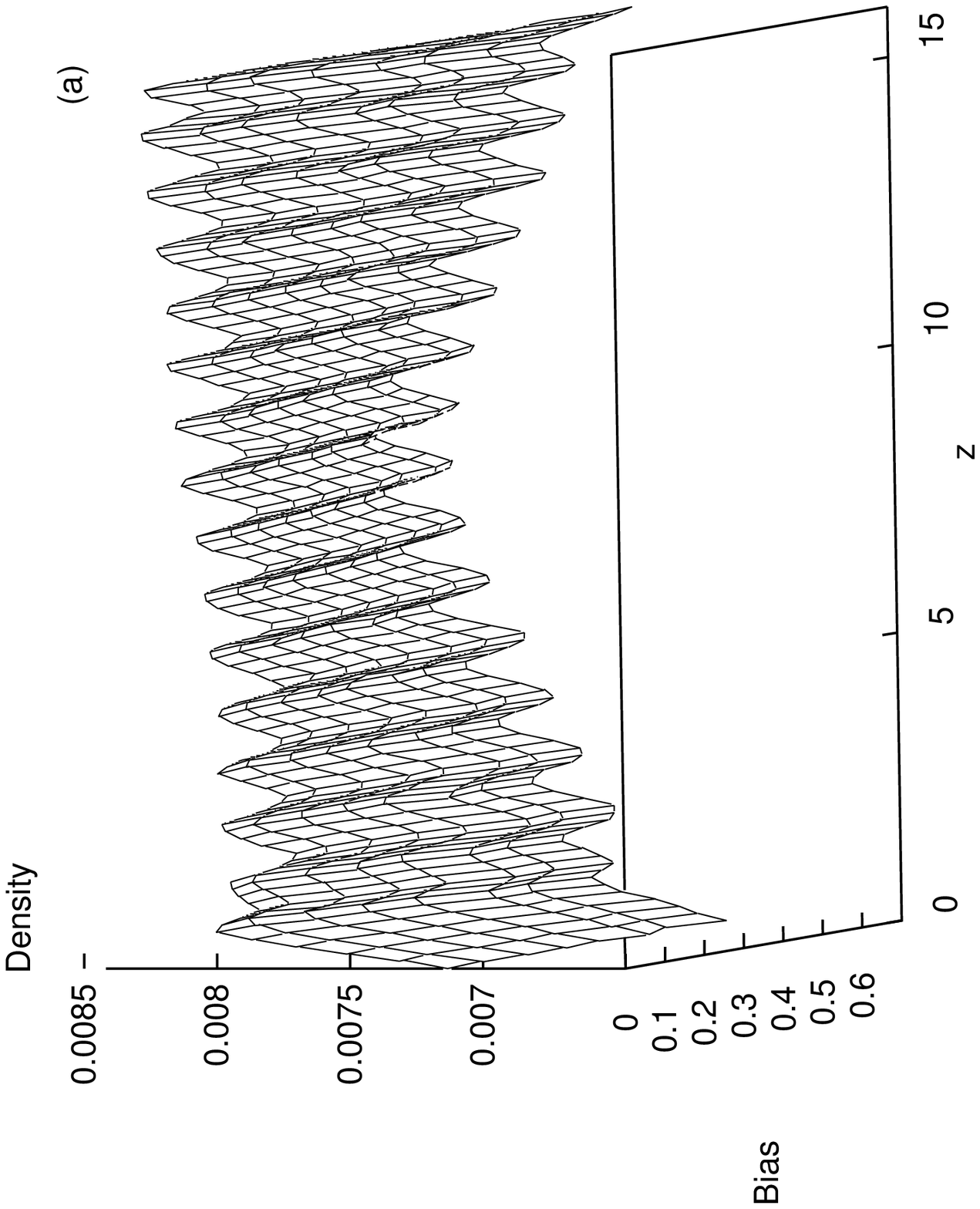}
\includegraphics{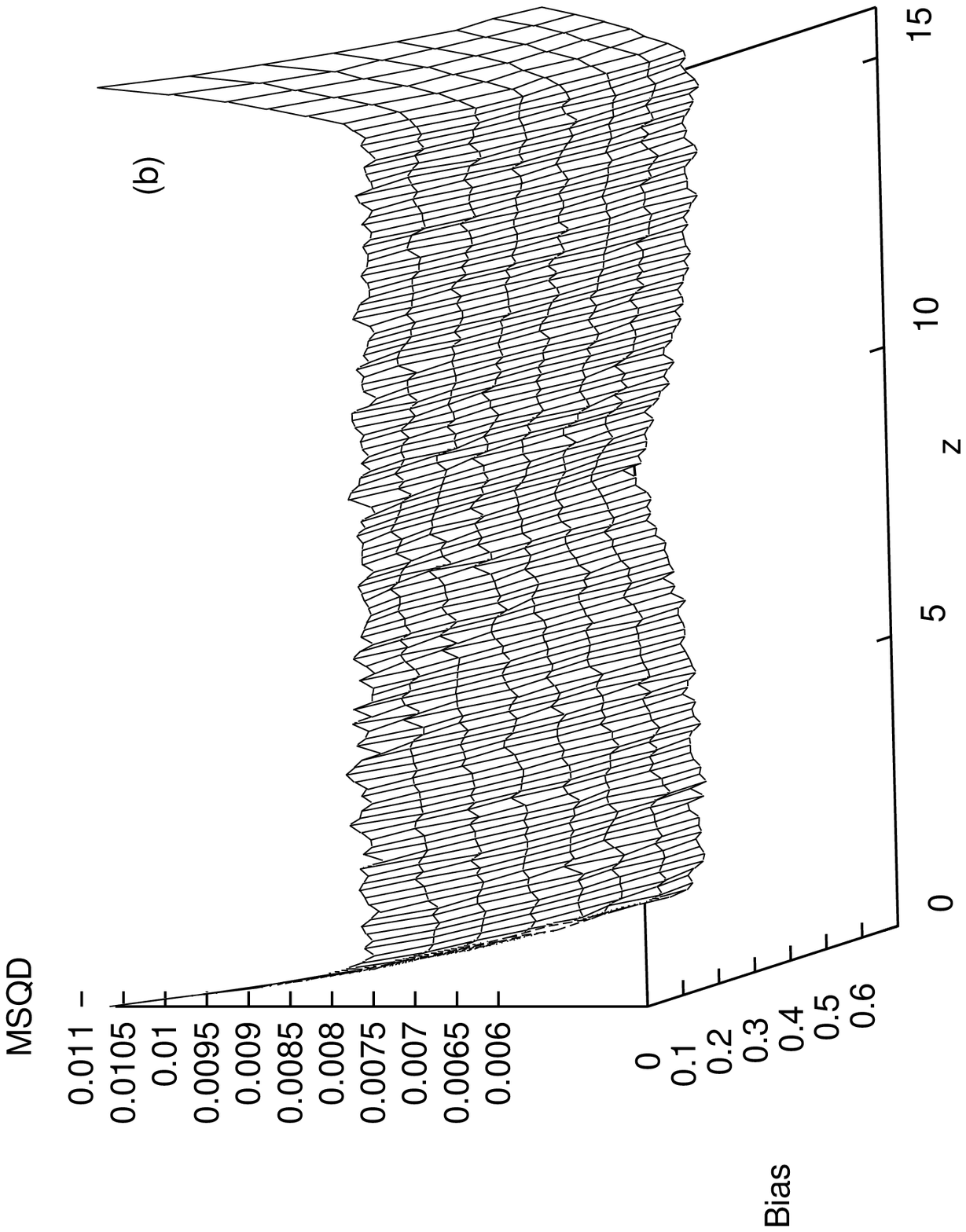}
\vskip 5.5cm
\includegraphics{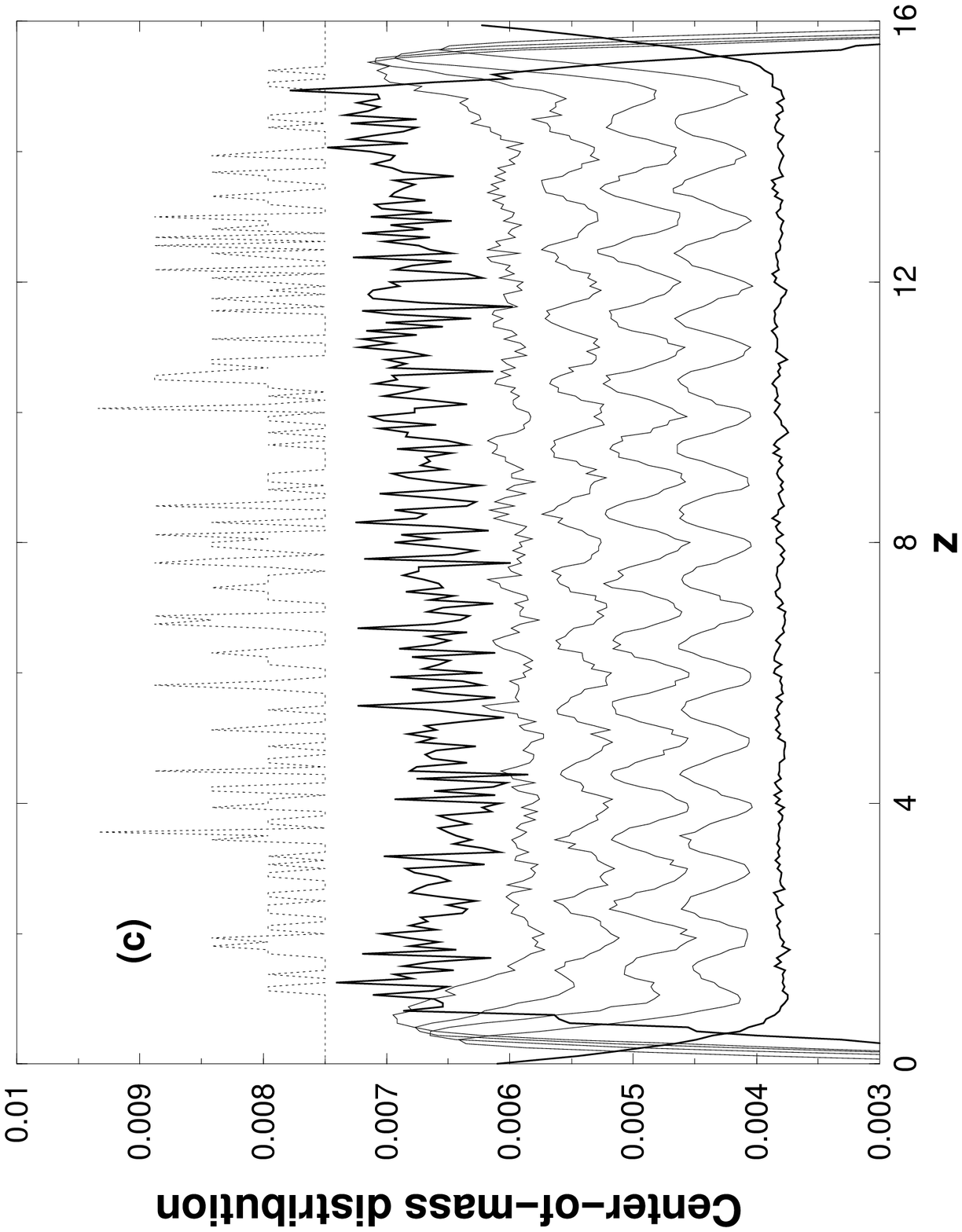}
\includegraphics{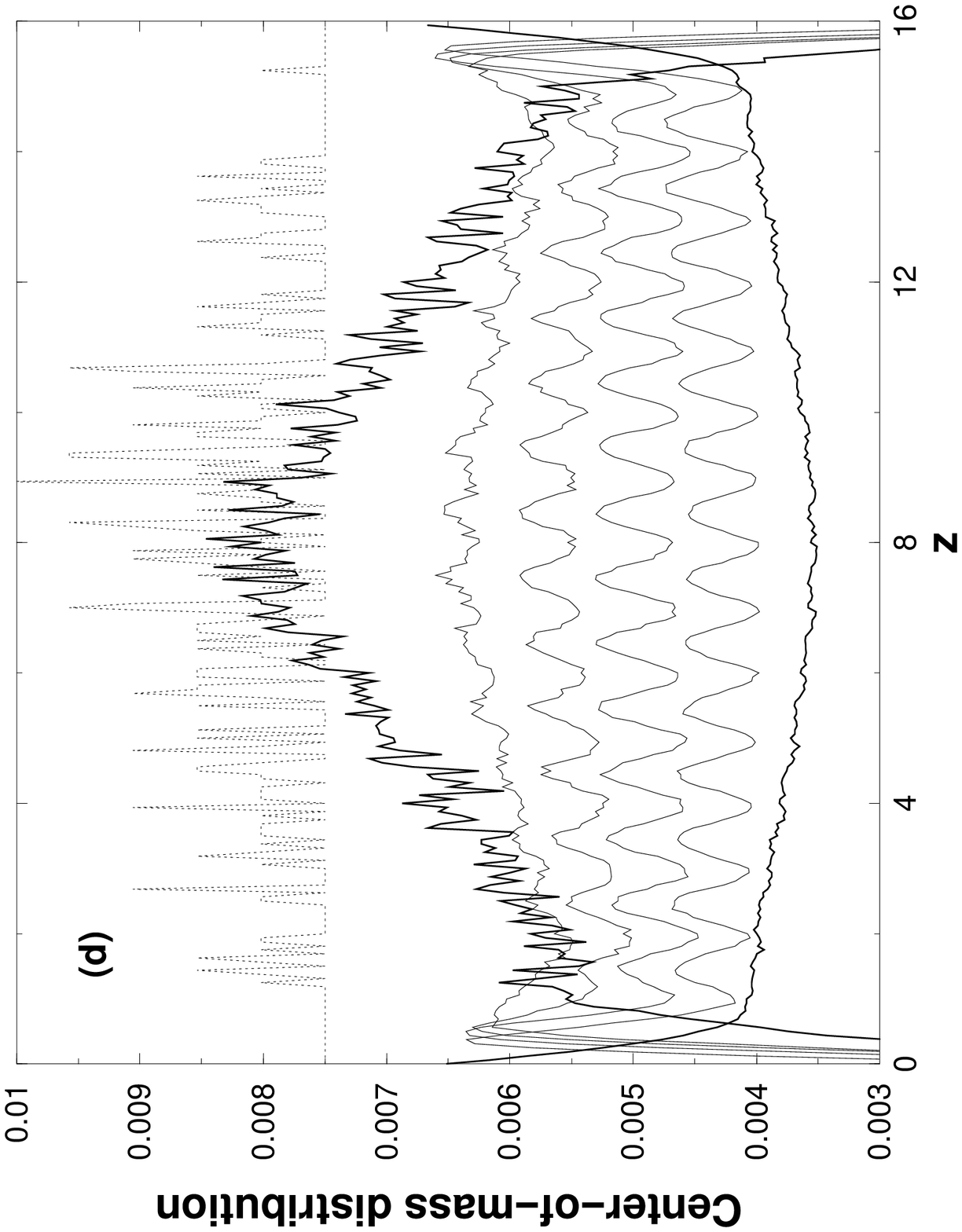}
\vskip 5.5cm
\caption{\label{1} (a) Distribution of the total monomer density across a slit
with $D=16$ at various rates of shear (bias).  (b) Average mean square
displacement after $1$ MCS. (c) Center-of-mass distribution of chains
with $l = 1, 2, 3, 4, 5, 10\; \mbox{and}\; 20$ for $B = 0$. Each 
curve is shifted from the previous one along the $y$-axis at $0.0005$ for better
visibility.  (d) The same as in (c) but for $B = 0.7$.
The width of the slit is $Z_{max} = 16,\;c = 0.5$ and $J/kT = 1$.}
\end{figure}

Most of the simulational results, discussed in the preceding subsections,
have been carried out for a sufficiently strong energy bond, $J/kT = 7$, which
is equivalent to a rather low temperature of the system. The average chain length
at $J/kT = 7$ thereby varies with density within the interval $40 \le L 
\le 70$ where the flexibility of the chains ensures that their conformations 
correspond to well shaped polymer coils.
It is interesting to check whether a change in the mean size of the chains 
$L$ in some way
affects the reorganization of the polydisperse system under shear flow. If
one reduces the ratio of bond to thermal energy $J/k_BT$, as mentioned in the
previous section, Eq.(\ref{L}), the average contour chain length decreases exponentially
fast. In the present study we change $J/kT$ from $7$ down to $1$ whereby $L$
drops from $\approx 40$ to $2.5$. As shown below this leads to dramatic changes
in the EP solution. The profiles along the $z$ axis for this case
of very short chains are shown in Fig. \ref{1}.
The oscillations in the the total density, Fig. \ref{1}, suggest a transition
of the system into an ordered state of nematic liquid crystal with an easy
direction parallel to the walls. Evidently the presence of hard walls acts as
an external ordering field on the chains. The system is dominated by monomers,
dimers and other very short species which behave largely like stiff rods
aligning themselves parallel to the walls. Indeed, Figs. \ref{1}c, \ref{1}d
demonstrate that this ordering is most pronounced for dimers, trimers and
tetramers whereas neither single monomers nor chains with length $l \ge 5$
participate in the ordering. Thus both single monomers, which lack any
anisotropy in shape, and longer chains with conformations of coils rather 
than stiff rods are insensitive to the ordering influence of the walls.
  
The influence of growing shear rate on the system with $L = 2.5$ is similar
to that in the case of $L \approx 40$ too. 
From Figs. \ref{1}c, \ref{1}d one may conclude
that the longer chains, which are otherwise uniformly distributed, now
pull closer to the middle of the slit where the flow velocity is zero.
This tendency starts with the $4-$ and $5$-mers already and is very
clearly seen for the $10$-mers (there are only few $20$-mers in the system
 at $J/k_BT = 1$ and their statistics is therefore rather poor). 
This effect more than compensates the developing shallow minimum in single 
monomer concentration in the middle of the box.

%% file: end.tex
\section{Concluding Remarks}
The present simulational study of the impact of shear flow on EP in a slab
reveals a number of interesting features:
\begin{itemize}
\item{The average chain length in a system of EP decreases steadily with
growing shear rate}
\item{The polymer coil is gradually stretched along the flow direction
as the shear is increased} 
\item{The MWD in a dilute solution of EP changes qualitatively when
sufficiently strong shear rate is imposed from Schwartz- to a mean-field
like exponential distribution function.}
\item{The shear rate introduces inhomogeneity in the system of EP, confined
in a slit: the monomer density, the diffusion coefficient and the concentration
of macromolecules with different lengths develop characteristic profiles 
perpendicular to the walls}
\item{The width of the depletion layer near the wall for long chains 
grows with increasing shear rate in agreement with recent 
EWIF studies}
\end{itemize}
Another interesting phenomenon - an ordering transition in a system
dominated by the shorter and
stiffer chains is found to take place upon {\em heating} of the system of EP 
with the result that chain length is reduced.
In this case shear flow is observed
to enhance the degree of ordering in the system. 

Our observations show that the relaxation of a system of EP from a state
of rest to that of steady state flow is a slow process which requires
long time intervals of investigation, probably rendering Monte Carlo 
simulational 
methods probably more appropriate than Molecular Dynamics.

We should like to note, however, that the shear rates studied in the
present work are limited to low and moderate values since stochastic jumps
along and against the external field may be biased by means of the Boltzmann
factor in a MC procedure within the framework of $100\%$ at most. We
therefore expect that at even higher shear rates the influence of
flow on EP properties might be more dramatic. Clearly, additional work
and adequate alternative methods are still needed to reach comprehensive
understanding of the problem.

\section{Acknowledgments}
This research has been supported by the National Science Foundation, Grant 
No. INT-9304562 and No. DMR-9727714, and by the Bulgarian National Foundation
for Science and Research under Grant No. X-644/1996. J.~W. acknowledges 
support by EPSRC under Grant GR/K56233 and is indebted for hospitality
in the Center for Simulational Physics at the University of Georgia.

%% file: draft.bbl
\begin{references}

\bibitem{ref}
A broad class of surfactants forming long flexible (worm-like) aggregates,
called {\em giant micelles} (GM), consists of a polydisperse solution of
differently long macromolecules which constantly recombine with each other,
or break into smaller fractions\cite{CC}. In contrast to GM in {\em living
polymers} (LP) equilibrium polymerization proceeds by means of a fixed amount
of initiator, which activates one of the chain ends\cite{Greer1} so that
single monomers may attach or break at the polymer end only. IN many aspects
the behavior of GM and LP in equilibrium is very similar, e.g. both are
characterized by an exponential Molecular Weight Distribution (MWD).
\bibitem{CC}
M. E. Cates and S. J. Candau, J. Phys. Cond. Matt. {\bf 2}, 6869(1990).
\bibitem{Greer1}
S.~C.~Greer, Adv. Chem. Phys. {\bf 96}, 261(1996).
\bibitem{WMC2}
J. P. Wittmer, A. Milchev, and M. E. Cates, J. Chem. Phys. {\bf 109}, 834(1998).
\bibitem{K}
M.~Kr\"{o}ger, and R. Makhloufi, Phys. Rev. E {\bf 53}, 2531(1996).
\bibitem{Greer}
S. C. Greer, Comput. Mater. Sci. {\bf 4}, 334(1995).
\bibitem{Cates}
M. E. Cates, Of Micelles and Many-Layered Vesicles, in {\em Theoretical
Challenges in the Dynamics of Complex Fluids}, Ed. T. McLeish, NATO ASI 
Series, pp. 257-283, and references therein.
\bibitem{MWL1}
A. Milchev, J. Wittmer and D.~P.~Landau, (submitted for publication).
\bibitem{AMYRDL}
A.~Milchev, Y.~Rouault, and D.~P.~Landau, Phys. Rev. E {\bf 56}, 1946(1997).
\bibitem{MilchevPotts}
A.~Milchev, Polymer {\bf 34}, 362 (1993);
\bibitem{ML}
A.~Milchev and D.~P.~Landau,J. Chem. Phys. E {\bf 104}, 9161(1997).
\bibitem{Berret}
J.~F.~Berret, D.~C.~Roux, G.~Porte and P.~Lindner, Europhys. Lett.
{\bf 25}, 521(1994).
\bibitem{Schmitt}
V.~Schmitt, F.~Lequeux, A.~Pousse and D.~Roux, Langmuir, {\bf 10}, 955(1994).
\bibitem{Makhloufi}
R.~Makhloufi, J.~P.~Decruppe, A.~Ait-Ali and R.~Cressely, Europhys. Lett.
{\bf 32}, 253(1995).
\bibitem{Furo}
I.~Furo and B.~Halle, Phys. Rev. E {\bf 51}, 466(1995).
\bibitem{Gelbart}
S.~Q.~Wang, W.~M.~Gelbart and A.~Ben-Shaul, J. Phys. Chem. {\bf 94},
2219(1990).
\bibitem{Wang}
S.~Q.~Wang, Macromolecules, {\bf 24}, 3004(1991).
\bibitem{Spenley}
N.~A.~Spenley, M.~E.~Cates and T.~C.~McLeish, Phys. Rev. Lett. {\bf 71},
939(1993).
\bibitem{Rondel}
F.~Rondelez, D.~Ausserre, and H.~Hervet, Ann. Rev. Phys. Chem. {\bf 38},
3056(1987).
\bibitem{deG}
P.~G.~de Gennes, J. Chem. Phys. {\bf 60}, 5030(1974).
\bibitem{Aubert}
J.~H.~Aubert and M.~Tirrell, J. Chem. Phys. {\bf 72}, 2694(1980).
\bibitem{Onuki}
A.~Onuki, J. Phys. Soc. Japan, {\bf 54}, 3656(1985); A.~Onuki,
Phys. Rev. Lett. {\bf 62}, 2472(1989).
\bibitem{Biller}
P.~Biller and F.~Petruccione, J. Non-Newton. Fluid Mech. {\bf 25},
347(1987).
\bibitem{Goh}
C.~J.~Goh, J.~D.~Atkinson, and N.~Phan-Thien, J. Chem. Phys. {\bf 82},
988(1984).
\bibitem{Diaz}
F.~G.~Diaz, J.~Garcia de la Torre, and J.~J.~Freire, Polymer {\bf 30},
259(1989).
\bibitem{Duering}
E.~Duering and Y.~Rabin, Macromolecules {\bf 23}, 2232(1990).
\bibitem{WPB}
J.~Wittmer, W.~Paul, and K.~Binder, Macromolecules, {\bf 25}, 7211(1992).
\bibitem{Gerroff}
I. Gerroff, A. Milchev, W. Paul, and K. Binder, J. Phys. Chem. {\bf 98}, 6526(1993).
\bibitem{MPB}
A.~Milchev, W.~Paul, and K.~Binder, J. Phys. Chem. {\bf 99}, 4786(1993).
\bibitem{J}
J. Kriegl, Adv. Polymer Sci. {6}, 170(1969).
\bibitem{Zimm}
B.~H.~Zimm, J. Chem. Phys. {\bf 24}, 269(1956).
\bibitem{Yet}
A.~Yethiraj, and C.~K.~Hall, Macromolecules, {\bf 23}, 1865(1990).
\bibitem{PMB}
R.~Pandey, A.~MIlchev, and K.~Binder, Macromolecules, {\bf 30}, 1194(1997).
\bibitem{Aussere2}
D.~Aussere, J.~Edwards, J.~Lecourtier, H.~Hervet, and F.~Rondelez,
Europhys. Lett. {\bf 14}, 33(1991).
\end{references}
